\documentclass{aa}  
\usepackage{graphicx}
\usepackage[figuresright]{rotating}
\usepackage{txfonts}
\begin{document}
   \title{VV Pup in a low state: secondary-star irradiation or stellar activity?\thanks{Based on
   data collected on the ESO VLT within the program 272.D-5044(A)}}

   \subtitle{}

   \author{Elena Mason
          \inst{1}
          \and
           Steve B. Howell \inst{2}
           \and
	  Travis Barman \inst{3}
	  \and 
	  Paula Szkody \inst{4}
	  \and 
          Dayal Wickramasinghe \inst{5}
          }

   \offprints{E. Mason}

   \institute{European Southern Observatory (ESO)
             Alonso de Cordova 3107, Vitacura, Santiago, CL\\
              \email{emason@eso.org}
              \and NOAO, 950 N. Cherry Ave., Tucson, AZ USA \\
             \email{howell@noao.edu}
         \and Lowell Observatory, Planetary Research Center, 1400 W. Mars Hill Rd., Flagstaff, AZ USA \\
             \email{barman@lowell.edu}
	 \and Dept. of Astronomy, University of Washington, Seattle, WA USA\\
	     \email{szkody@astro.washington.edu}
         \and Australian National University, Australia\\
             \email{dayal@maths.anu.edu.au}}

   \date{Oct 2007; ??? }

 
\abstract{}{Emission lines in polars show complex profiles with multiple components that are typically ascribed to the accretion stream, threading region, accretion spot, and the irradiated secondary-star. In low-state polars the fractional contribution by the accretion stream, and the accretion spot is greatly reduced offering an opportunity to study the effect of the secondary-star irradiation or stellar activity. We observed VV Pup during an exceptional low-state to study and constrain the properties of the line-forming regions and to search for evidence of chromospheric activity and/or irradiation.}
{We obtained phase-resolved optical spectra at the ESO VLT+FORS1 with the aim of analyzing the emission line profile and radial velocity as a function of the orbital period.  We also tailored irradiated secondary-star models to compare the predicted and the observed emission lines and to establish the nature of the line-forming regions.} {Our observations and data analysis, when combined with models of the irradiated secondary-star, show that, 
while the weak low ionization metal lines (FeI and MgI) may be  consistent with irradiation processes, the dominant Balmer H emission lines, as well as NaI and HeI, cannot be reproduced by the irradiated secondary-star models. We favor the secondary-star chromospheric activity as the main forming region and cause of the observed H, NaI and He emission lines, though a threading region very close to the L1 point cannot be excluded. } {} 


   
   \maketitle
%

\section{Introduction}

Different  types  of cataclysmic  variables  (CVs,  i.e. dwarf  novae,
polars, etc)  in different states (outburst or  quiescence) have shown
Balmer emission lines forming  close to the secondary-star.  Depending
on  the  system,  plausible  origins  of  these  emission lines  have  been
identified as  the irradiated  secondary hemisphere (e.g.   Steeghs et
al. 2001; Araujo-Betancor  et al. 2003; Thoroughgood et  al. 2005), or
the accretion stream (e.g. Cowley  et al. 1982; Mukai 1988; Schwope et
al.   1997).  More  recently, it  has  been proposed  that the  Balmer
emission  lines  in  low-state  polars could  be  the  signature  of
chromosphere activity on the secondary-stars in AM Her, ST LMi, and EF
Eri (Kafka et  al.  2006, 2007, Howell et  al.  2006b).  This scenario
is  particularly intriguing  if  we consider  that V471~Tau  H$\alpha$
emission line,     previously      thought     to     be      caused     by
irradiation\footnote{Young   et  al.    (1988)  use   the   word  {\it
fluorescence}  in the  broader context  of conversion  of  high energy
photons to low energy ones.}  (Young et al. 1988), is best explained by
stellar activity (Rottler et al.  2002).

In  order to  understand the  source  and formation  mechanism of  the
observed Balmer emission lines in  low-state polars, we  analyzed time-resolved optical spectra of VV~Pup  obtained during a low-state.  The
spectra  were secured  at  the  ESO VLT+FORS1  and  have been  already
presented in Mason  et al. (2007), where we focused on the white dwarf magnetic field signatures and, in particular, the first detection of the Zeeman absorptions. Here we present  in Sec.~\ref{uno} the data sample, in Sec.~\ref{due} our
radial  velocity study  and the  emission line  profile  analysis.  In   Sec.~\ref{tre} the  irradiated
atmosphere model tailored for the VV~Pup system is compared to the observations. Sec.\ref{quattro} summarizes our results and conclusions.  


\section{Data overview \& major spectroscopic features}
\label{uno}
 
\begin{figure}
\centering
\includegraphics[angle=0,width=8.0cm]{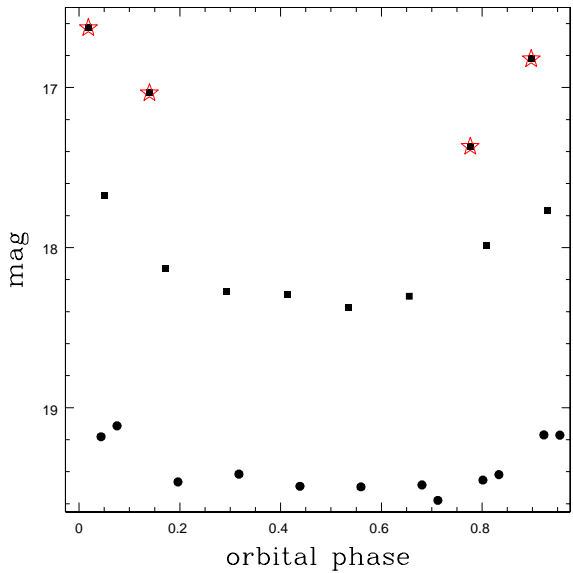}
\caption{Light curve extracted from the two data set by convolving each spectrum with a V-Johnson filter. Circles represents the data set of March 16. Squares represent the data set of March 22. Stars are for the last 4 frames in the March 22 data set which correspond to an increased mass transfer rate or to a flare episode.}
\label{fig1}
\end{figure}

The data were secured at the ESO VLT+FORS1 on 2004 March 16 and 22 (hereafter epoch 1 and epoch 2, respectively). The observation strategy and data reduction have already been presented in Mason et al. (2007), therefore we direct the reader to that paper for the details. 
Here we briefly present the two time series data sets and their observational characteristics. Each time series consists of 12 spectra and covers 1.26 orbital periods. 
The two series of spectra are very similar to each other as they both show strong cyclotron humps from the main and the secondary magnetic poles (see Figure 1 in Mason et al. 2007) and similar emission lines (mainly HI, NaI and HeI, but also see below). The cyclotron hump emission modulates the system light curve (Fig.~\ref{fig1}): VV Pup is brighter at the time of the secondary inferior conjunction when the main accreting pole on the white dwarf is in view, and fainter at the time of the secondary superior conjunction, when we face the white dwarf secondary magnetic pole. 

However, the two time series data sets do have differences as they correspond to two different brightness states of the binary system. VV Pup was about 1 mag brighter during the epoch 2 observations than at the time of the epoch 1 observations.  Moreover, epoch 2 observations showed a sudden increase in brightness in the last 4 spectra of our time series which corresponds to the phase range 0.80-1.15. This is a possible signature of a flare and/or a sudden increase of the mass transfer rate (accretion burst), as we observe also an increase in flux of the emission lines  (see Sec.~\ref{due}). 

\begin{figure*}
\includegraphics[angle=270,width=18.0cm]{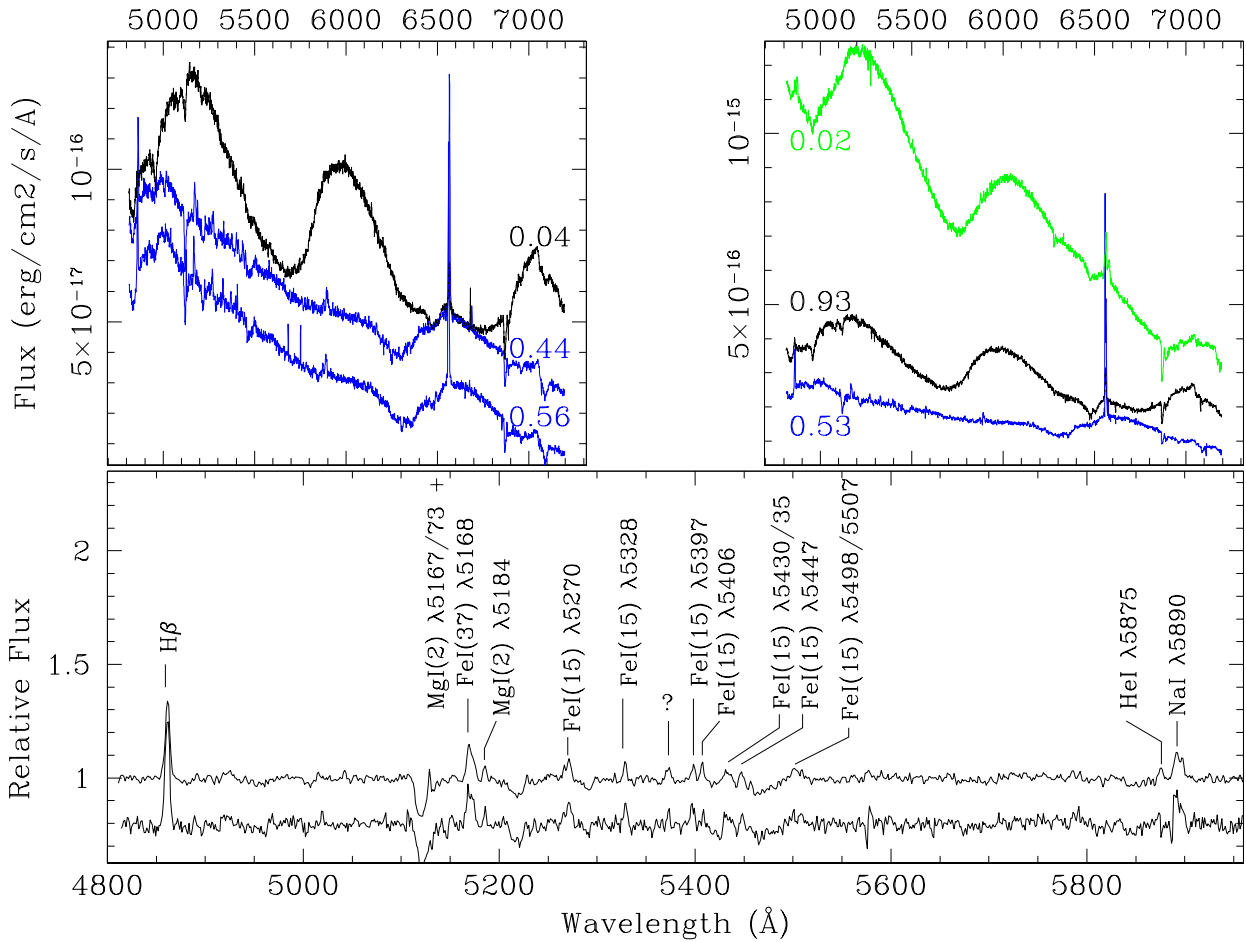}
\caption{Top panels: bright- (black) and faint- (blue) phase spectra of each data set, epoch 1 on the left and epoch 2 on the right. The numbers near each spectrum indicate their orbital phase. In the top left panel the spectrum at orbital phase 0.56 has been shifted vertically by the constant -0.2e-16 erg cm$^2$ sec$^{-1}$. In the top right panel the green spectrum is the bright-phase spectrum observed during the accretion burst. Bottom panel: a zoomed-in view of the 4800-5960\AA \ region with the identification of the low ionization energy metal lines. The spectra in the bottom panel have been normalized. The top spectrum (in the bottom panels) is the average of the spectra at orbital phases 0.44 and 0.56 (properly shifted before co-addition) of epoch 1 data set; while the bottom spectrum is that at orbital phase 0.53 in epoch 2 data set. The bottom spectrum in the bottom panel has been vertically shifted by the constat -0.25.}
\label{fig2}
\end{figure*}

Figure~\ref{fig2} shows the bright- and faint-phase spectrum in each series. Despite the different system brightness, corresponding spectra at the two epochs are similar. In particular, the bright-phase spectra (orbital phase $\sim$0) are dominated by cyclotron humps from the 32 MG magnetic pole and just weak H$\alpha$ and H$\beta$ emission lines. The faint-phase spectra (orbital phase $\sim$0.5) show cyclotron humps from the 55 MG magnetic pole and relatively strong emission lines from H, and low ionization energy metal lines. We recognize emission lines from the MgI (2) triplet (the strongest among the metal lines), the FeI (15) multiplet and the NaI D doublet. The latter is flanked, on the blue side, by the HeI (11) emission $\lambda$5876 (Fig.~\ref{fig2}'s bottom panel).

\section{Emission line properties: observational evidence}
\label{due}
\subsection{Balmer emission lines}

In this section we analyze in detail the emission line properties (profile characteristics, radial velocities, and fluxes) to investigate their origin. We start with the Balmer emission lines which have higher S/N and, therefore, provide more robust measurements. 

\begin{figure*}
\includegraphics[angle=0,width=9.0cm]{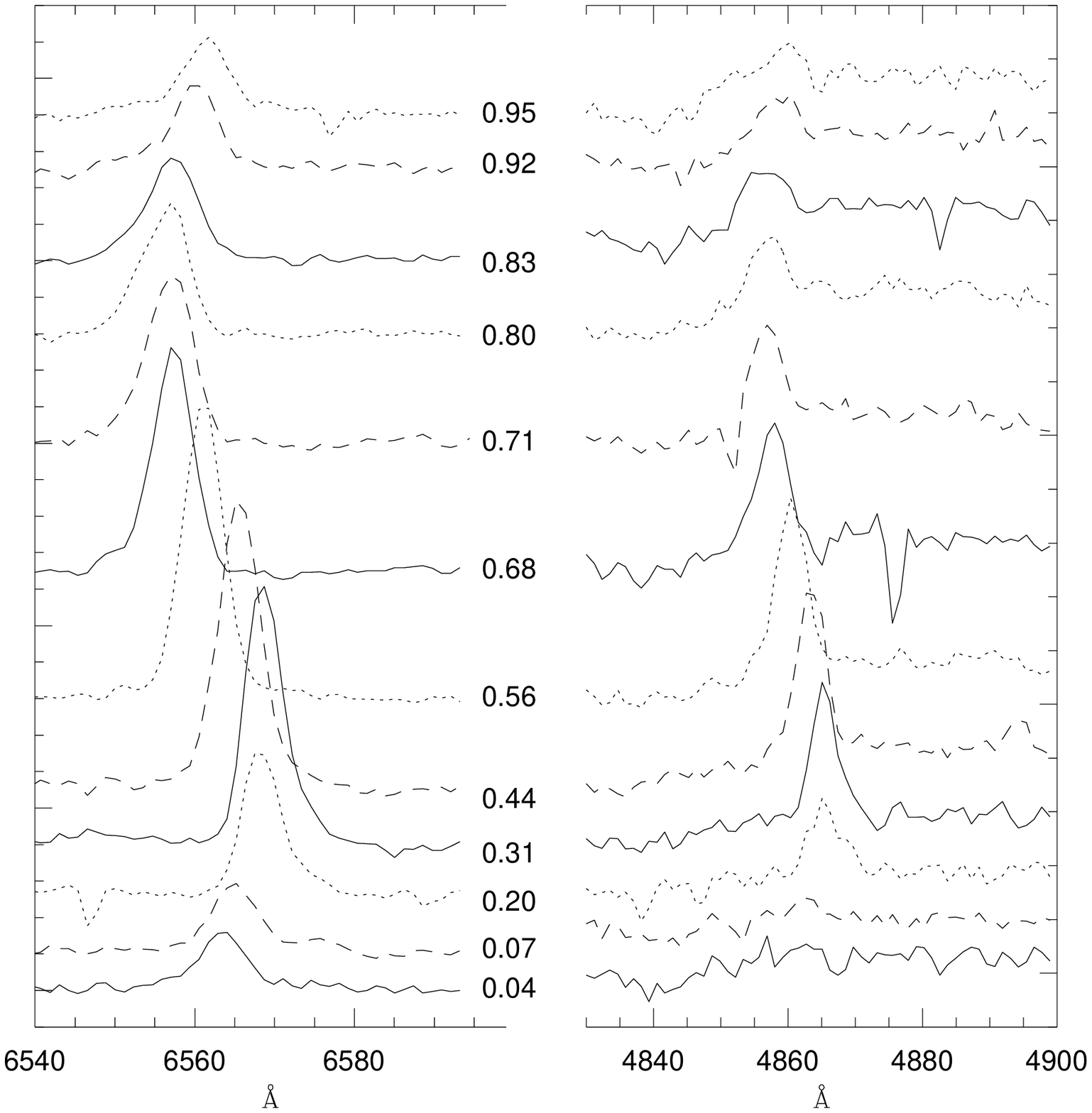}
\includegraphics[angle=0,width=9.0cm]{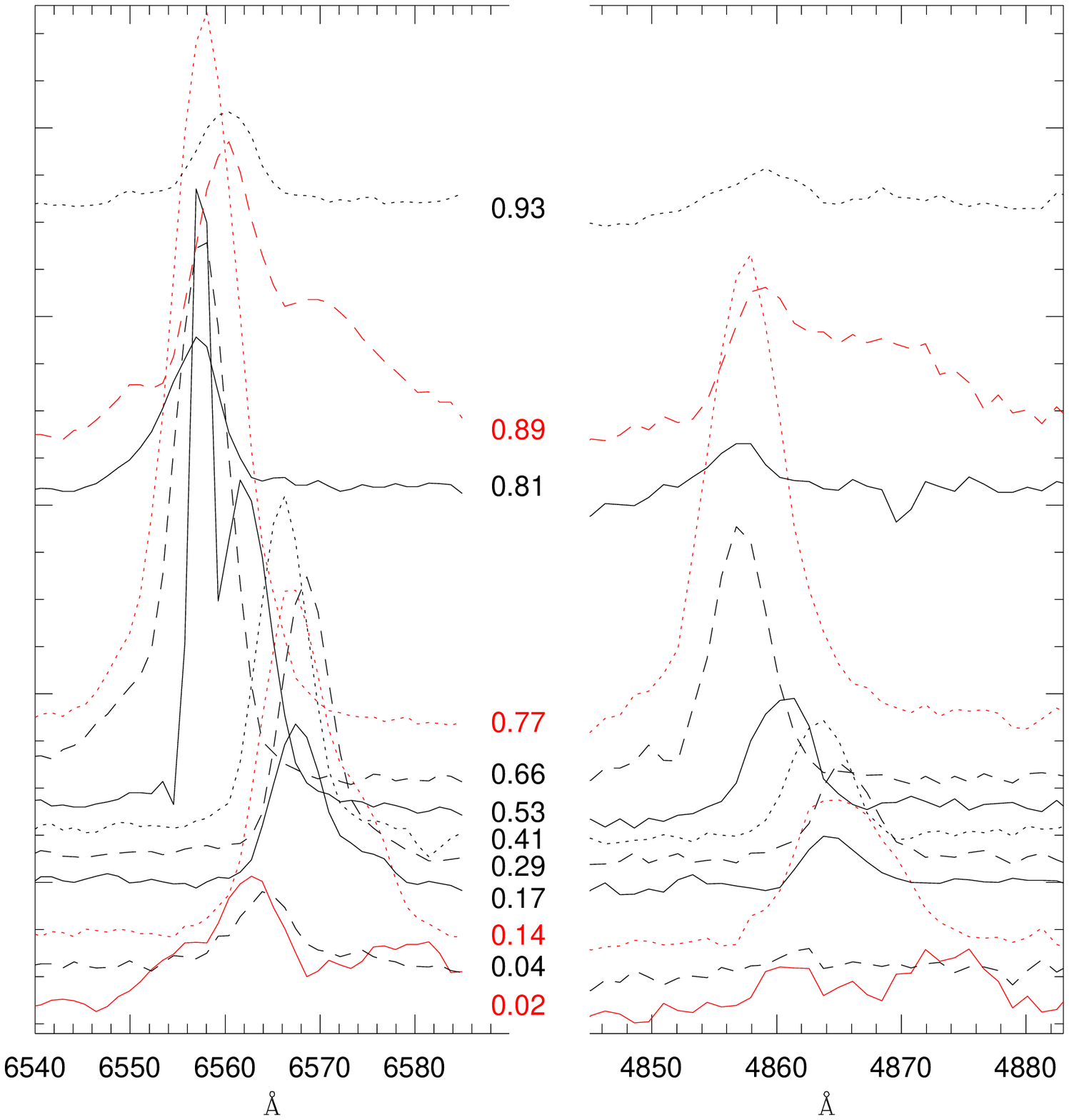}
\caption{Balmer emission line profiles for epoch 1 (left panel) and epoch 2 (right panel). Different line styles are just for clarity, while the lines in red color correspond to the four epoch 2 flaring/bursting spectra.}
\label{fig3}
\end{figure*}

\begin{figure*}
 \includegraphics[angle=270,width=18cm]{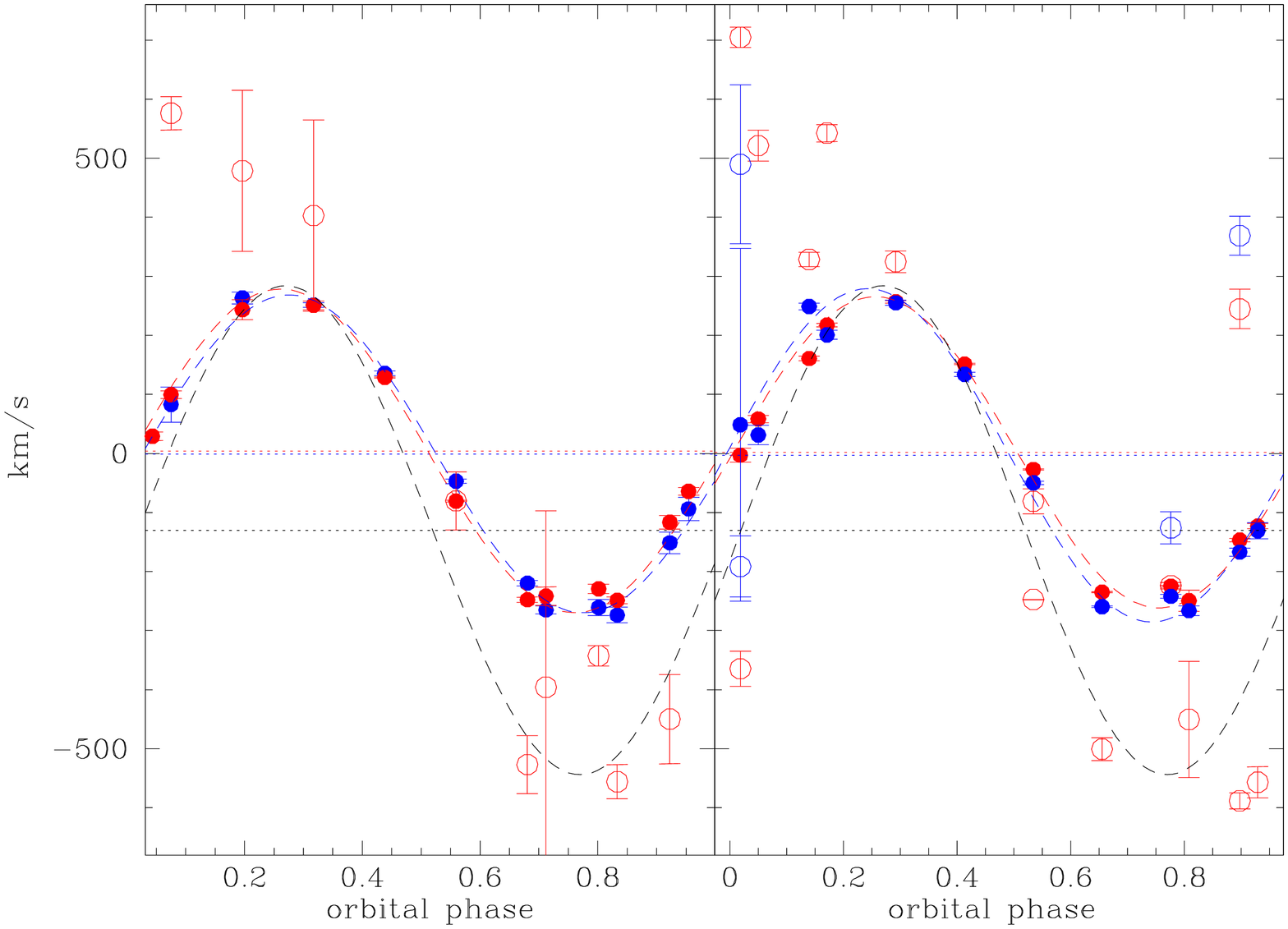}
\caption{Radial velocity measurements (left epoch 1, right epoch 2) of the Balmer lines H$\alpha$ (red solid circles) and H$\beta$ (blue solid circles) and their best fit (red and blue dashed lines).  Open circles represent the secondary-satellite component (same color convention), while the dotted horizontal lines represent the $\gamma$ velocity as determined by the sinusoidal fit. In black we report the best fit radial velocity curve determined by Howell et al. (2006a, sinusoidal dashed line) and its $\gamma$ velocity (horizontal dotted line).}
\label{fig4}
\end{figure*}

\begin{figure*}
 \includegraphics[angle=0,width=9.0cm]{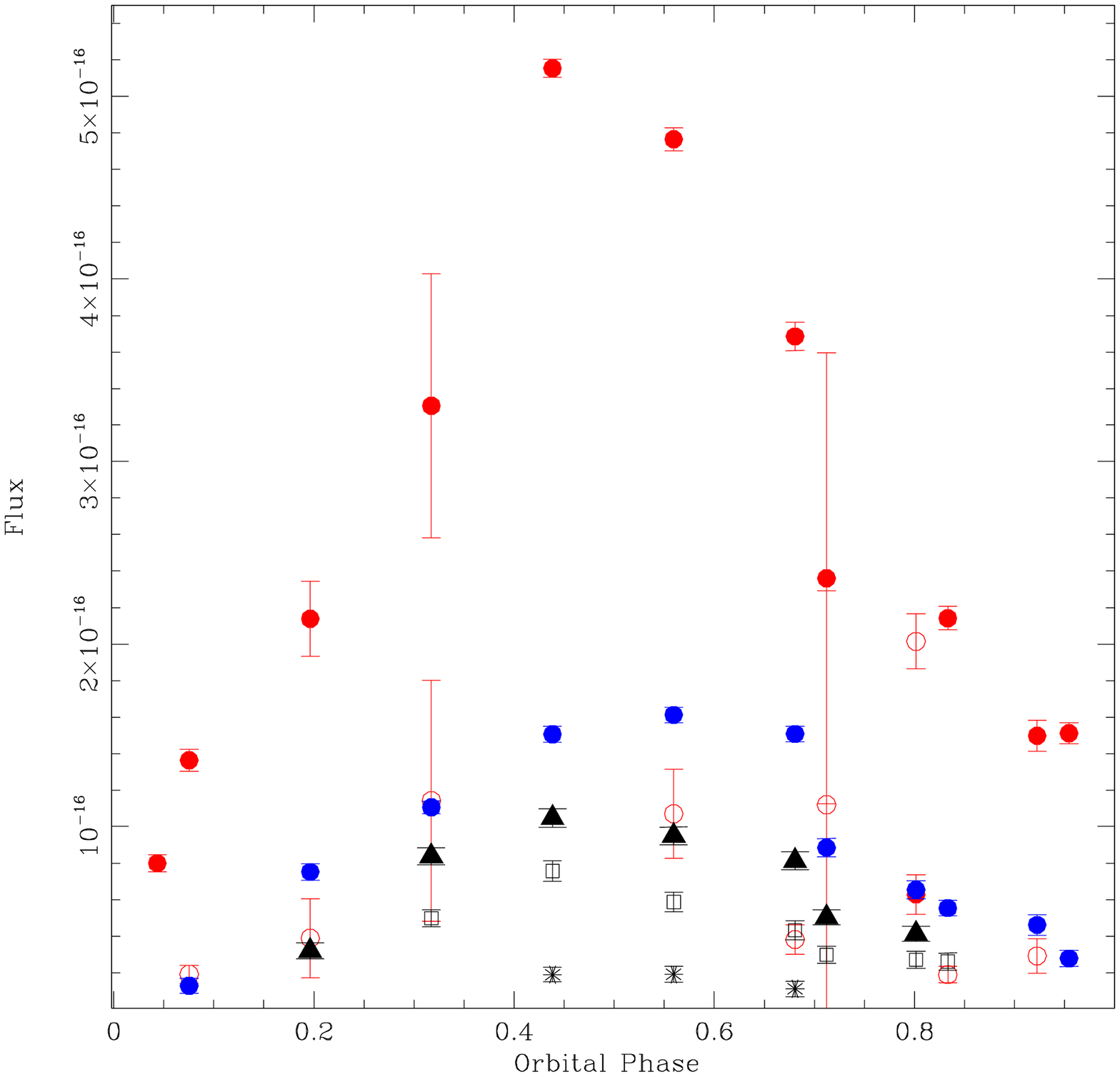}
 \includegraphics[angle=0,width=9.0cm]{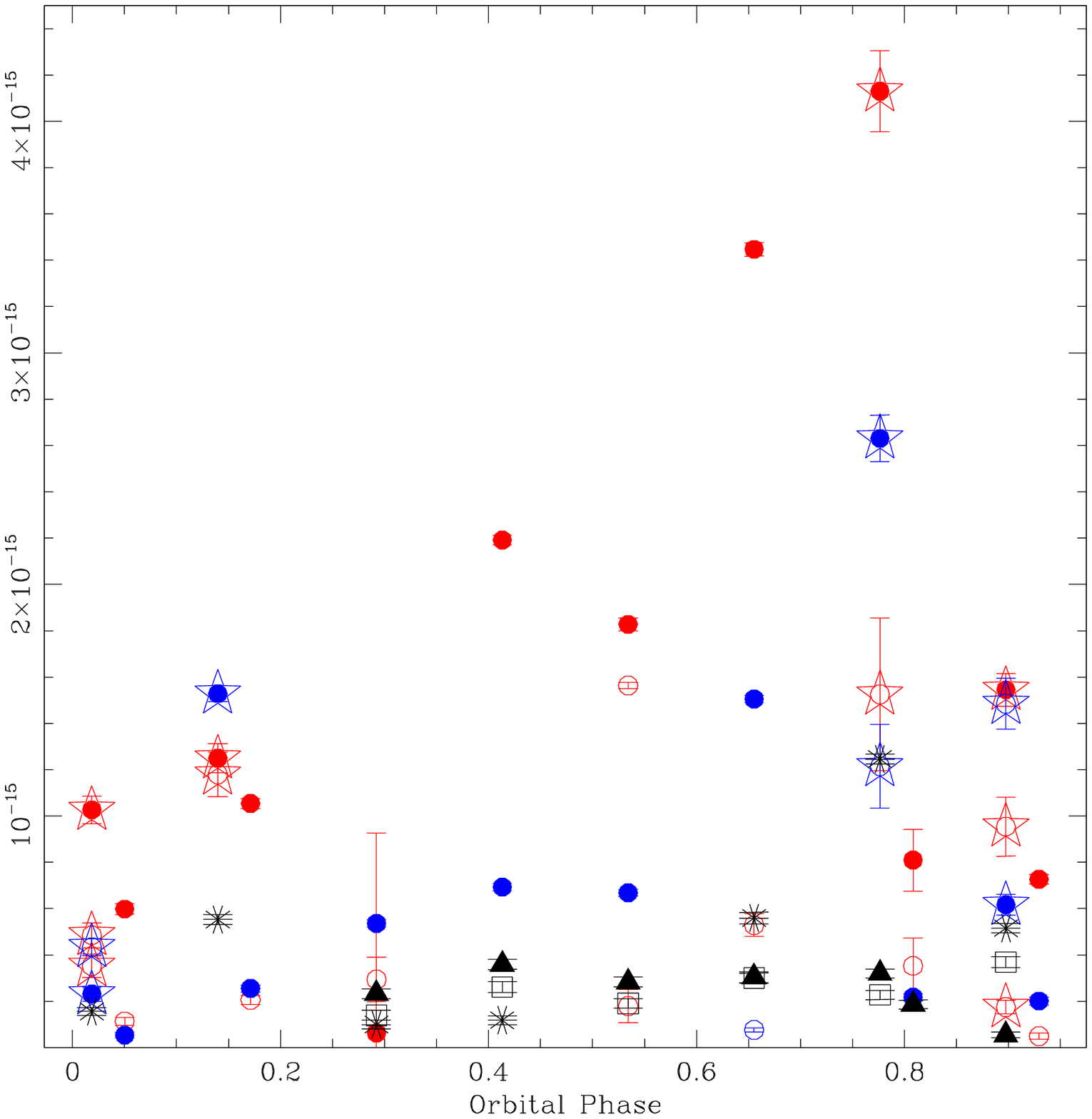}
\caption{Emission line flux modulation versus the orbital period. Epoch 1 data are in the left panel, while epoch 2 data are displayed in the right panel. In both panels, red solid circles are for the H$\alpha$ main-narrow component, while red outlined circles are for any of the additional components found via Gaussian fit. Similarly, blue solid and empty circles refer to the H$\beta$ emission lines. Black triangles are for the MgI line, empty black squares for the NaI doublet, and black asterisks for the HeI line. In the right panel, the starred symbols mark the Balmer emission lines in the 4 flaring/bursting spectra of epoch 2.}
\label{fig5}
\end{figure*}

Figure~\ref{fig3} shows the profile of both the H$\alpha$ and H$\beta$ emission lines at the two epochs. The line intensity varies with the orbital phase being stronger around phase 0.5 (see also Fig.~\ref{fig5}), while the line profile depends on the epoch. The emission lines appear as mostly single Gaussian (or single component) emission lines in the epoch 1 spectra, while clearly multiple components (2 or 3) are present in epoch 2 spectra. In order to assess the significance of multiple components at both epochs we fit each line of each spectrum both with a single and multiple (2 to 3) Gaussians. We then compared the significance of the multiple fit with respect to the single Gaussian fit with the F-test. We eventually adopted a multiple Gaussian fit whenever its probability exceeded $\geq$3.5$\sigma$. The result is that in epoch 1 the H$\alpha$ emission lines are better fit by a double Gaussian in 75\% of the cases (i.e. 9/12 spectra), while the H$\beta$ line is better fit by a single Gaussian in all but one case (where there was no detection of the H$\beta$ emission line, 11/12 spectra or 92\%). 
In epoch 2 a multiple Gaussian fit (2 or 3) is needed in 93\% of the cases (10/12 spectra) for the H$\alpha$ emission line and in 25\% of the cases (3/12 spectra) for the H$\beta$ line. 
The radial velocities corresponding to each component are shown in Fig.~\ref{fig4} where solid circles represent the main narrow  component (FWHMs are of the order of 2-3\AA), while the open circles are for the secondary, weaker  component\footnote{Note that the FWHM of the secondary component in the epoch 1 spectra is not larger than 2\AA, while, in epoch 2 spectra it varies (with the orbital phase) between 2 and 6\AA.}. Similar weaker components have also been identified in AM~Her (Kafka et al. 2006), EF~Eri (Howell et al. 2006b) and ST~LMi (Kafka et al. 2007) and called ``satellite'' lines. Figure~\ref{fig4} shows that the main narrow component has a clean sinusoidal pattern and is in phase with the secondary-star.  We fitted the H$\alpha$ and the H$\beta$ main narrow components at each epoch. The best fit parameters are shown in Fig.~\ref{fig4} as well as in Table~\ref{tab1}. 

\begin{table}
\centering
\caption{The best fit radial velocity curve solutions for the Balmer (narrow component) and the ``non-Balmer'' lines at the two epochs. }
 \begin{tabular}{ccccc}
epoch & line ID & R/B crossing & systemic velocity & K \\
 & & & & \\
1 & H$\alpha$ & 0.489$\pm$0.001 & 5$\pm$2 & 277$\pm$2 \\
1 & H$\beta$ & 0.473$\pm$0.003 & -1$\pm$3 & 267$\pm$3 \\
2 & H$\alpha$ & 0.497$\pm$0.001 & 4$\pm$1 & 267$\pm$1 \\
2 & H$\beta$ & 0.509$\pm$0.001 & -3$\pm$2 & 283$\pm$2 \\
& & & &\\
1 & FeI 5270 & 0.469$\pm$0.023 & -36$\pm$35 & 326$\pm$19 \\
1 & FeI 5328 & 0.493$\pm$0.028 & 63$\pm$41 & 313$\pm$22 \\
1 & MgI 5170$^\dagger$ & 0.447$\pm$0.007 & -43$\pm$11 & 341$\pm$9 \\
1 & NaI 5893 & 0.476$\pm$0.016 & 2$\pm$21 & 317$\pm$14 \\
1 & HeI 5875 & 0.440$\pm$0.135 & -38$\pm$190 & 250$\pm$66 \\
& & & &\\
2 & FeI 5270 & 0.498$\pm$0.020 & 7$\pm$40 & 371$\pm$38 \\
2 & FeI 5328 & 0.471$\pm$0.082 & -1$\pm$132 & 289$\pm$46 \\
2 & MgI 5170 & 0.478$\pm$0.010 & -12$\pm$12 & 329$\pm$11 \\
2 & NaI 5193$^{\dagger\dagger}$ & 0.468$\pm$0.015 & -11$\pm$20 & 323$\pm$16 \\
2 & HeI 5875 & 0.568$\pm$0.008 & 6$\pm$6 & 178$\pm$5 \\
& & & & \\
 \end{tabular}
$^\dagger$ Barycenter of the blend MgI(2)$\lambda\lambda$5167,5173.\\
$^{\dagger\dagger}$ Barycenter of the blend NaI D $\lambda\lambda$5190,5196.
\label{tab1}
\end{table}

The best fit parameters in Table~\ref{tab1} appear quite similar to each other. From here on, however, we regard the epoch 1  H$\beta$ radial velocity solution as the most reliable for showing the line location within the binary system as it is not biased  at any phase by a secondary emission component. 
Because of the phasing of the radial velocity curve, it is tempting to associate the line with the secondary-star and conclude that it mirrors the secondary-star orbital motion. However, Howell et al. (2006a) have determined the secondary-star radial velocity curve by measuring the NaI absorption at 2.067 micron. A comparison of our primary, narrow emission line radial velocity fit with that by Howell et al. (2006a), shows that we agree in the phase of the red to blue crossing (i.e. the secondary superior conjunction), but not in K amplitude, nor in the $\gamma$ velocity determinations. Our Keplerian velocity is significantly smaller than that given by Howell et al. (2006a) while our $\gamma$ velocity is $\sim$140 km/s more positive. We can attempt to reconcile the measurements by applying the inverse K velocity correction to our determination on the hypothesis that the Balmer emission lines originate from the heated face of the secondary-star. Assuming the most extreme unrealistic case of the emission line forming just at the L1 point (Horne and Schneider 1989) we derive K$\sim$408 km/sec, which is only 6 km/sec smaller than the Howell et al. (2006a) result. The more realistic assumption of the emission line rising from about half the secondary-star surface produces K=323 km/sec which is significantly smaller (by$\sim$90 km/sec) than Howell et al. (2006a) result. Moreover, this scenario still leaves unexplained the significant difference in the $\gamma$ velocities.

The study carried out by Mukai (1988) has shown that in the case of (high rate) mass transfer we can expect emission lines phased with the secondary-star either from the horizontal part of the accretion stream (before the coupling region where the gas gets funneled by the white dwarf magnetic field lines), or in the rising part of the funneled gas just after (or at) the threading region (emission lines from the gas falling onto the white dwarf along the magnetic field lines will necessarily be phased with the primary-star). 

In the case of emission lines from the horizontal stream, we should observe K amplitudes which are larger than K2, the real secondary-star Keplerian velocity. The red-to-blue crossing phase and $\gamma$ velocity should be similar to those of the real secondary-star radial velocity curve. 
Hence we can exclude emission from the horizontal stream before it is funneled in the magnetic field lines as the cause of the narrow main H emission line source. 

In the specific case of the VV~Pup low-state, it is difficult to imagine a line-forming region just on the rising part of the gas that has been funneled along the magnetic field lines, as one would also expect that emission lines would form in the falling part and the horizontal stream (if present), as well as in the impact region. We do not observe either one.  We could expect emission lines from the threading region provided that this happens to be very close to the secondary-star. This might actually be the case for short orbital period systems and low-state systems (the magnetosphere radius is inversely proportional to the mass transfer rate to the power of -2/11, hence it moves outward and far from the primary star during low-states. See Warner 1995 for a review and Mukai 1988 for details.). 
However, the high velocities ($\sim$300 km/sec) and the chaotic motion of the gas in the threading region are quite inconsistent with our observed FWHMs and possibly also (but not necessarily) with our derived $\gamma$ velocity.  

Hence, we favor the hypothesis of an extended atmosphere and/or chromospheric prominence-loops from the  secondary-star. Prominence-loops can easily be out of the binary orbital plane thus explaining (in analogy with Mukai consideration of the funneled accretion stream) both the small K amplitude, the relatively small line width  and the (more positive) $\gamma$ velocity. In the hypothesis that these prominence-loops are somehow triggered or enhanced by the white dwarf magnetic field (Howell et al. 2006b, see also Uchida and Sakurai 1985 Figure 1) it is reasonable that they would not be visible during the secondary-star inferior conjunction.   
With regard to this, we show in Fig.~\ref{fig5} the modulation of the emission line flux versus the orbital period. We identify a well defined bell shape distribution for epoch 1 measurements (left panel, red and blue symbols for H$\alpha$ and H$\beta$, respectively) as expected for lines which form on just one side of a rotating star. 
The more scattered distribution for epoch 2 measurements (Fig.~\ref{fig5} right panel, red and blue symbols for H$\alpha$ and H$\beta$, respectively) is explained both by the bias introduced by the 4 flaring/bursting spectra (starred symbols) as well as by the overall line profile which shows a more extendend (/multiple) line-forming region(s) than in epoch 1 spectra; i.e. multiple components.

\subsection{The non-Balmer emission lines}

We can attempt a similar analysis for the weaker non-Balmer emission lines, namely the metal emission lines (MgI and FeI), NaI and HeI. As they are weak and often blended (e.g. the MgI 5167, 5173 and, possibly, FeI 5168 are all blended together), and  given the instrumental resolution, we limit our measurements to the line flux barycenter and the integrated flux. We also limit our measurements to the two FeI lines $\lambda$5270 and $\lambda$5328, which appear to be the strongest among multiplet~(15). 

\begin{figure*}
 \includegraphics[angle=0,width=9.0cm]{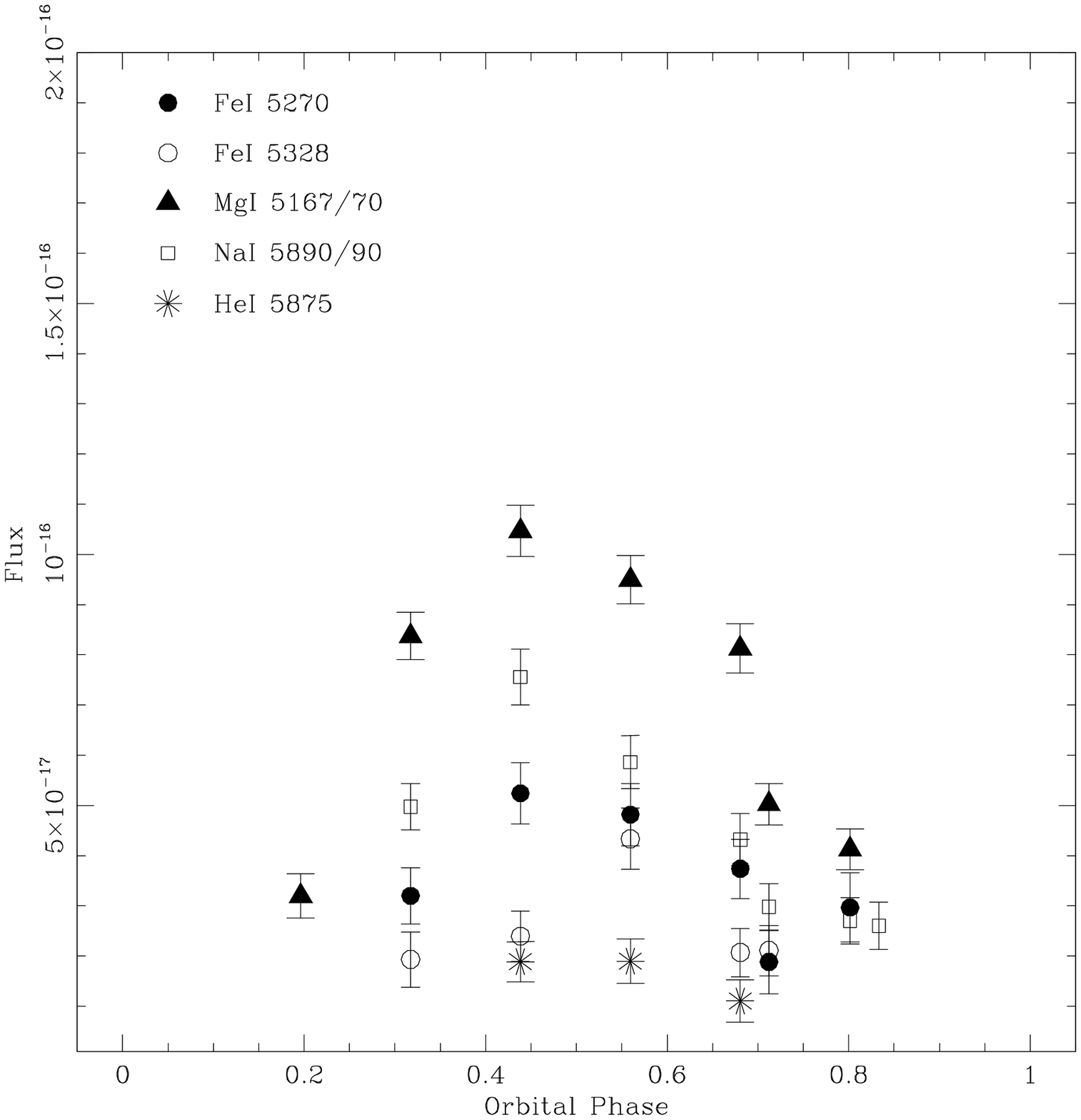}
 \includegraphics[angle=0,width=9.0cm]{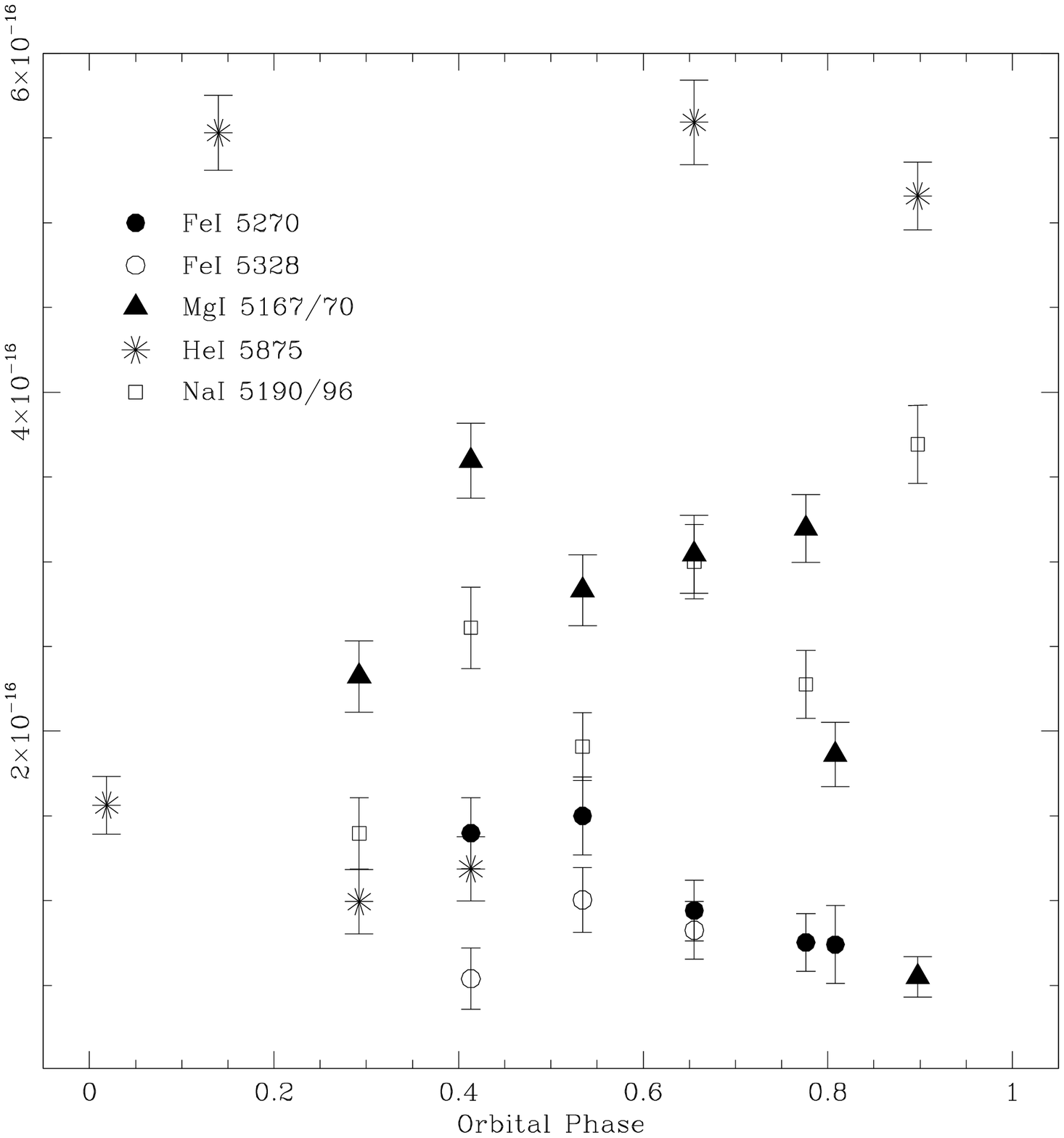}
\caption{Zoomed-in view of the emission line flux distribution versus orbital phase. The left and the right panels display the data of epoch 1 and 2, respectively. Different symbols are for different emission lines (notation on the figure itself). Note that in the case of the HeI emission the data point at orbital phase 0.78 has not been plotted for clarity, being a factor of two brighter than the brightest points displayed. }
\label{fig6}
\end{figure*}

We plot in both Fig.~\ref{fig5} and Fig.~\ref{fig6} the flux measurements of the HeI and the low ionization energy emission lines. Figure~\ref{fig5}  shows the flux modulation of the weak lines in comparison to the Balmer lines, while Fig.~\ref{fig6}  shows only the weak, non-Balmer lines in a rescaled plot for clarity. Note that the FeI measures were not plotted in Fig.~\ref{fig5} (again for clarity). 

The figures show that the low ionization energy metal lines and the NaI and the HeI emission lines in epoch 1 mirror the Balmer line behavior and follow a bell shaped modulation throughout the orbital period. 
In epoch 2, however, though we are still seeing the emission lines generally strongest around secondary superior conjunction, the flux distribution is more scattered. 
The measured line flux is at least double with respect to epoch 1 for all the lines, implying that they are all affected by the brighter state recorded in epoch 2. 
We note that the HeI emission line is different from the other lines in that it also becomes strong and visible in the bursting spectra at orbital phase 0.8-1.1, similar to the Balmer lines (Fig.~\ref{fig5}). 
 
\begin{figure*}
 \includegraphics[angle=0,width=9.0cm]{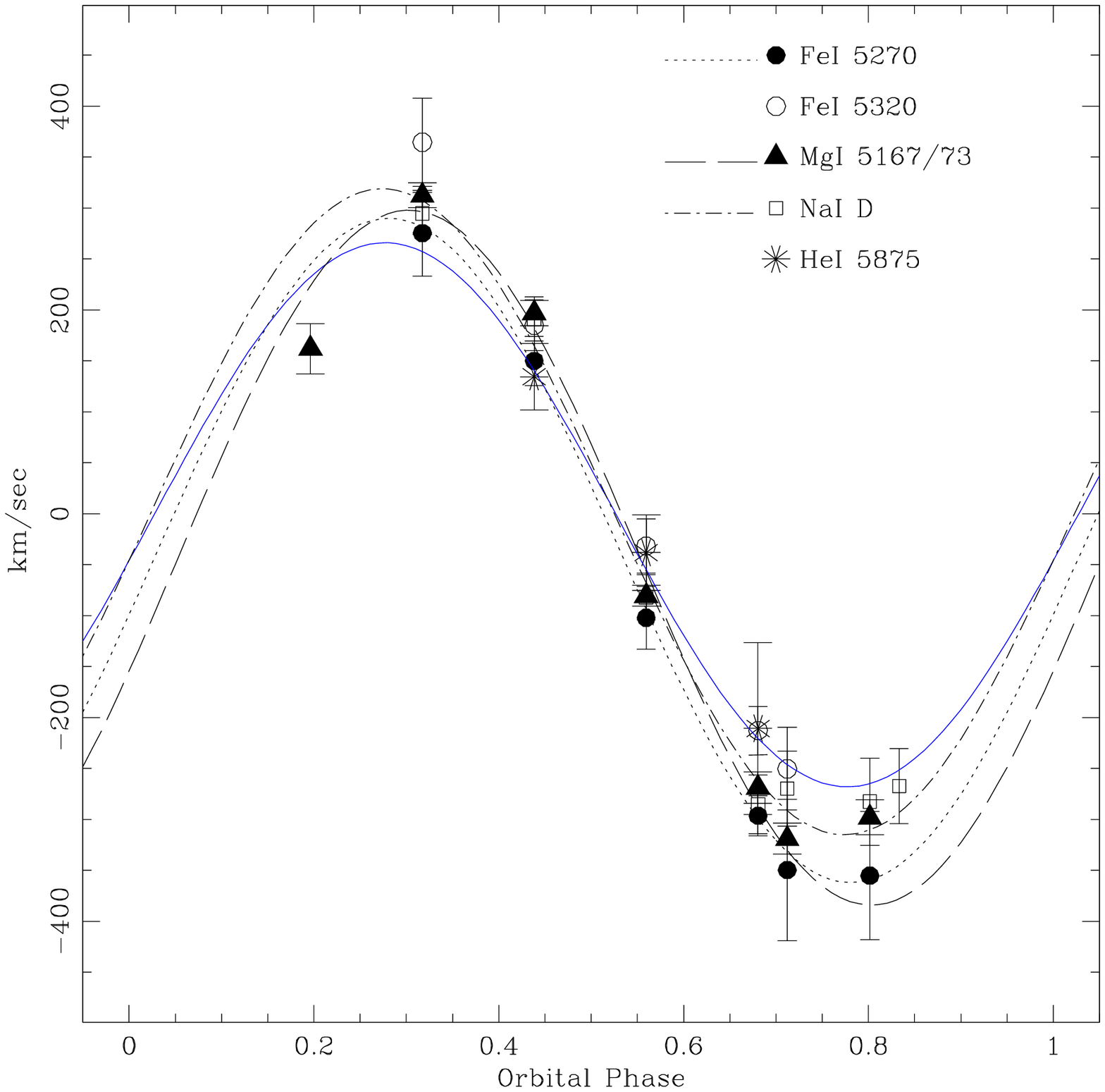}
 \includegraphics[angle=0,width=9.0cm]{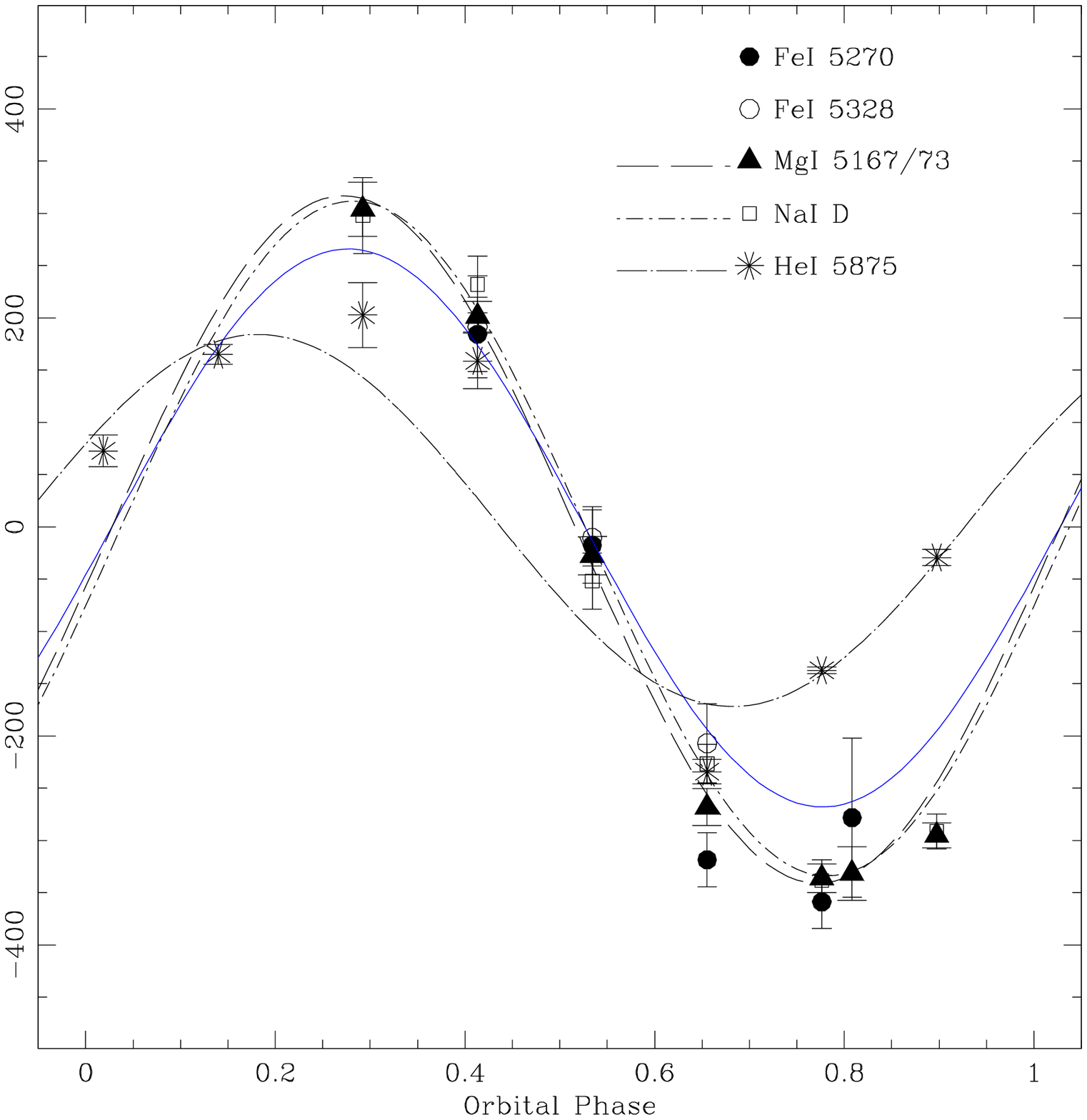}
\caption{Radial velocity measurements of the low ionization metal lines, (MgI and FeI) and the NaI and HeI emission lines. Left panel is for epoch 1 while the right panel is for epoch 2. Different symbols are for different emission lines and the notation is on the figure itself. Different line styles represent the best fit (as in Table~1) for the specified line. Note that not all best-fit radial velocity curves have been plotted for clarity. In both panels the blue solid line is the best fit determined for the H$\beta$ emission in epoch 1.}
\label{fig7}
\end{figure*}

The radial velocities for each of the measured lines and their best fits are displayed in Fig.~\ref{fig7} and Table~\ref{tab1}, respectively. 
These are characterized by larger errors. However, the overall velocity curves are very similar to those derived for the Balmer lines with the only difference being a possibly larger amplitude. We can imagine a line-forming region closer to the secondary-star and the L1 point in the case of the non-Balmer emission lines, or a temperature gradient in the line-forming region. 
We note that the HeI emission lines in epoch 2, again shows a very different behavior with respect to the other lines and epoch 1 measurements. A fit of the HeI lines shows a much smaller amplitude implying that the HeI lines, in the bursting spectra, form closer to the white dwarf but still on the secondary-star side of the binary system center of mass. 
The phase of the red to blue crossing is delayed as observed in the radial velocity measurements during high states (e.g. Cowley et al. 1982, Diaz and Steiner 1994, Schneider and  Young 1980). 
The HeI lines appear significantly stronger in the bursting spectra and this seems consistent with a picture where the emitting/threading region gets hotter due to a higher mass transfer rate and denser gas.

\section{The irradiated secondary-star model}
\label{tre}

Above we have  shown that the low-state emission  lines observed in VV
Pup  are consistent  with an  origin on  or near  the  secondary-star.
Three  scenarios for  their  production have  been  postulated in  the
literature:  threading region/accretion  stream  components, irradiation,  and  stellar
activity. Though some of our observational evidence may favor the threading region and/or the stellar activity scenario (see Sec.~\ref{due}), we could not completely rule out the irradiation case. 
We now want to test the irradiation hypothesis by producing synthetic models of the secondary-star irradiated by the white dwarf.

A  number  of papers,  both  about close  binaries  and  CVs, talk  about
irradiation  and present spectra  containing narrow  emission features
attributed   to  that  (see   Barman  et   al.   2004   and  reference
therein). Calculations of the energy needed for irradiation as well as
that supplied  by the white dwarf  have been presented  as well (e.g.,
Rottler et  al.  2002,  Schmidt et al.   1995).  Models  of irradiated
spectra have reached a  sophisticated level in Barman et  al. (2004), where it was shown that, in  the case of pre-CV  binaries having white  dwarfs of T$_{\sf
eff}  >$ 20000K,  the atmosphere  of the  secondary facing  the white
dwarf contains  a temperature inversion that leads  to strong emission
lines and  shallow absorption  features.  For lower  temperature white
dwarfs (T$_{\sf  eff} <$ 20000K),  it was shown that  the temperature
inversion in the secondary-star's upper atmosphere produces only weak,
narrow emission lines from low excitation ionized metals (mostly FeI).
Furthermore,  since the  temperature  inversion in  cooler systems  is
confined  to a  part of  the  atmosphere well  above the  photosphere,
photospheric absorption lines and bands are unaffected.

\begin{figure*}
   \centering         
   \includegraphics[angle=270,width=14cm]{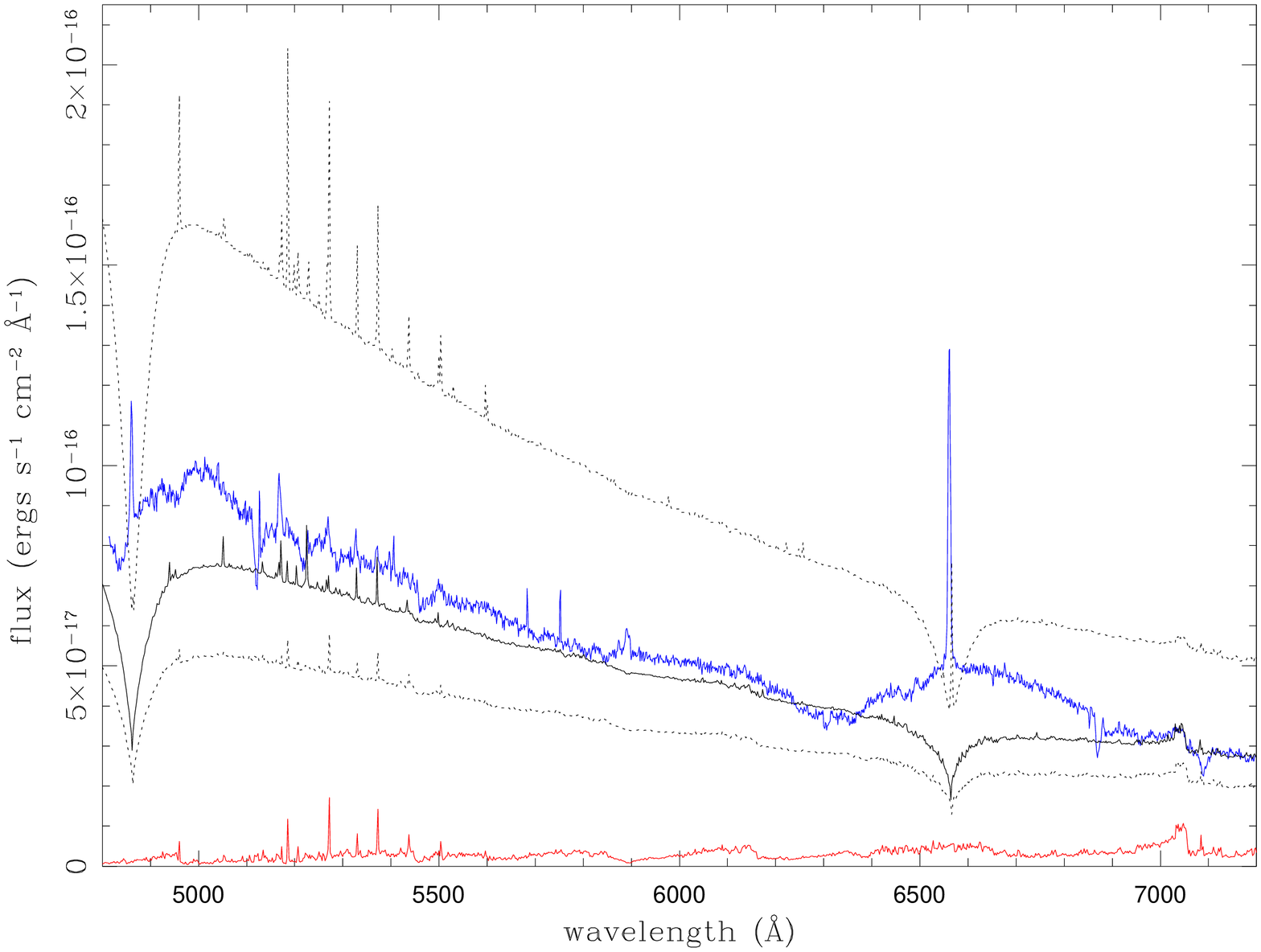}
   \caption{Models vs  observation comparison.  The  black spectra are
   the three  irradiation models  (white dwarf +  irradiated secondary
   star) described in Sec.~4. The dashed lines are for the LTE secondary
   atmosphere models, while  the solid line is for  the NLTE secondary
   atmosphere model.  The top spectrum  is for a 20000K  white dwarf,
   while the two  bottom ones are for a 11900K  white dwarf.  In blue
   color is the  VV~Pup epoch 1 faint phase spectrum. It is the average of the two spectra at phase 0.44 and 0.56 (see Fig.~\ref{fig2}). We also show, in red color, the NLTE irradiated secondary atmosphere model alone.  Flux  units refer  to  the  observed spectrum. The four models  have been arbitrarily scaled. However, the NLTE irradiated secondary-star model (red) and the NLTE irradiated secondary-star model  + the 11900K white dwarf (solid black) have been scaled by the same factor so that their relative flux and fractional contribution is preserved.. } 
              \label{fig8}
    \end{figure*}

Following the  model procedure outlined  in Barman et al.   (2004), the
white dwarf  in VV Pup was  modeled with a temperature  of 11900~K and
$\log g = 8.0 \pm  0.5$ (Araujo-Betancor et al.  2005).  The secondary-star  was modeled  assuming it to be  a normal,  solar metallicity  main
sequence  star of  spectral type  M7  (Howell et  al. 2006a),  $T_{\sf
eff}$=3000K and $\log g =  5.25$.  The secondary-star temperature and
gravity  were determined by  fitting the  2.067 $\mu$m NaI absorption
line profiles presented  in Howell et al.  (2006a). The  Na I line and
its local continuum  were assumed to be unaffected  by the white dwarf
flux as the $\sim$12000K white dwarf contributes less than 1\% of the
K-band flux in VV Pup.   Additionally, even if irradiation is present,
the 2.067 $\mu$m Na I absorption line will be  unaffected as it forms deep in the M
star photosphere, below the atmospheric layer at which any temperature
inversion would  occur.  Using these values and  the additional system
parameters  presented  in Howell  et  al.  (2006a),  we produced VV~Pup model
spectra (without cyclotron humps) by summing white dwarf  template spectra of different temperatures with
the irradiated  M7V LTE atmosphere model.   Figure~\ref{fig8} shows two
of these models  (black dashed lines): one having a  20000K white dwarf and the
other having  a 11900K  white dwarf. In  both cases the  white dwarf and  the irradiated
secondary spectra  have been rotationally  broadened to 25 and  200 km
sec$^{-1}$, respectively,  as appropriate for VV~Pup.  They also assume  
orbital phase 0.5 (a phase where the irradiation is expected to be the
largest)  and  orbital inclination  $i=90^o$. Since VV~Pup has an orbital
inclination $i\sim75^o$, the models line strength are an upper
limit for the expected irradiation.

Our secondary-star model  assumes a spherical  star, i.e.   the Roche
geometry is not accounted for.  Because of this, a small region of the
L1  area of  the secondary-star  atmosphere will  be slightly  closer
($\sim$ 0.1R$_2$,  see Fig. 6 in  Howell et al., 2000)  to the primary
and  have lower  gravity,  than in  our  spherical photosphere  model.
In order  to  explore the  consequences  of
neglecting the exact Roche  geometry, we computed several additional models artificially  moving the  secondary-star  closer  (but no closer than L1) to the 11900K white dwarf and arbitrarily lowering
its  surface gravity.   Each  of these  tests  produced only  slightly
greater   temperature  inversions,   without   altering  the   overall
irradiated spectrum appearance and our conclusions.

\begin{figure*}
\centering
\includegraphics[angle=0,width=14.0cm]{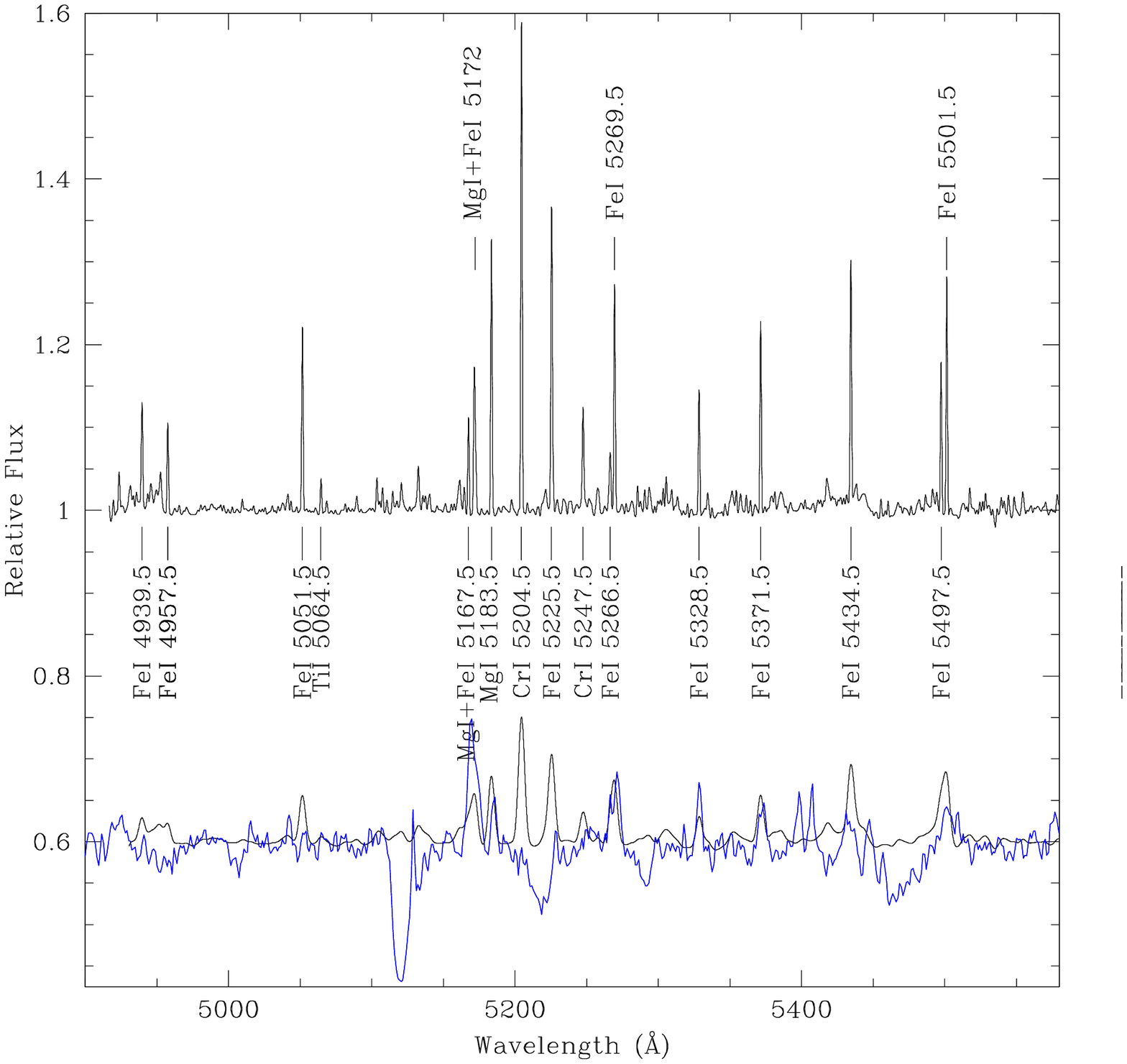}
\caption{Comparison of model and observation in the 4900-5600\AA \ region. In black color are the normalized model spectra at spectral resolution 0.5\AA \ (the resolution in the stellar atmosphere code, top) and 5.0\AA \ (about the resolution of our optical spectra, bottom). In blue color we plot the normalized faint phase spectrum of epoch 1, corresponding to the average of the two faint phase spectra at orbital phase 0.44 and 0.56. All the spectra have been normalized and shifted to a common reference wavelength. The observed blue spectrum and the low resolution model have been vertically shifted by the constant -0.4 for clarity.  }
\label{fig9}
\end{figure*}

Also, since the NLTE assumption is  more appropriate in the case of an
external radiation source, we also produced a model having a 11900K white dwarf 
and  a  NLTE irradiated  secondary  atmosphere  (black  solid line  in
Fig.~\ref{fig8}). This  latter model is the one  directly compared to
our  VV~Pup observations  (blue  solid line  in Fig.~\ref{fig8} and Fig.~\ref{fig9}).   We
might note  that the  use of LTE  secondary-star atmosphere  models in
place of the more proper NLTE one, does not alter the  conclusion as they predict
the same ``forest'' of low excitation energy metal lines, though with
reduced strengths. Figure~\ref{fig9} shows a zoomed-in view of the 4900-5600\AA \ region comparing the average normalized faint-phase epoch 1 spectrum (already shown in Fig.~\ref{fig2}) with the NLTE model, both at spectral resolution 0.5\AA \ and resolution matching our observations. Figure~\ref{fig9} shows that there is not a one-to-one matching of the observed and predicted lines and, in particular, that we do not observe CrI and TiI emissions. This might at least be explained in part by the model assumptions (see also below), as well as by the Zeeman absorption and the cyclotron humps, which ``bias'' our observed spectra.

Qualitatively  we  note that  the  irradiation  models predict  metal
emission lines mainly from FeI, TiI, MgI and CrI (i.e. low excitation,
singly  ionized  metals).  These  metal  lines  are  predicted to  get
stronger for  higher white  dwarf temperatures as  the increase  in UV
flux can  push the temperature  inversion deeper into the  cool star's
atmosphere.  Balmer,  HeI, and NaI  emission lines due  to irradiation
are not expected  in the case of a $\sim$12000K white  dwarf as in VV~Pup (See Fig.~\ref{fig8}), while only H and NaI emissions are expected for white dwarf temperatures $\geq$20000K.

Recent UV observations have pointed out the presence of a ``hot spot'' on the white dwarf surface (Szkody et al. 2006, Schwope et al. 2007) for low-state polars. This hot spot seems to have a temperature as high as 18000-24000 K and is present in all the low-state magnetic systems observed in the UV up to now (e.g. AM~Her G\"{a}nsicke et al. 2006). Araujo-Betancor et al. (2005) determined a slightly better fit to their UV spectra by including a second hotter component superimposed on the white dwarf model spectrum (either a power law or a second white dwarf spectrum), though they could not constrain its size and temperature. 
We have verified that model calculations which include a 25000K hot spot covering about 1\% of the white dwarf surface and superposed on the VV~Pup 11900K white dwarf, can indeed produce enhanced H emission together with a low ionization metal-line forest. 
A 25000K hot spot model does not noticeably affect the white dwarf optical continuum but produces too weak H emission lines to match our observations. In addition, the model predicts an inverse Balmer decrement, which is also in disagreement with our observations.  
We note that the relative strengths of the produced emission lines depend on several factors such as the  NLTE-effects (which alters the Boltzmann distribution of the population levels), the assumed abundances (in our case solar abundances), and atomic parameters (e.g. oscillator strengths and/or line centers). Still, within reasonable assumptions, these models do not seem capable of fully explaining the observations. At the same time, our data cannot independently provide evidence for or against the presence of a hot spot on the white dwarf. 
Only low-state UV observations similar to those of, e.g,  EF~Eri (Szkody et al. 2006, Schwope et al. 2007) can solve this issue.  We would like to point out that the EF~Eri optical spectrum is well fit by just a $\sim$10000K white dwarf model, which fails to fit the UV data (Schwope et al. 2007). 

\section{Summary and Conclusions}  
\label{quattro}

Phase-resolved optical spectroscopy of the polar VV~Pup during a low-state has shown the presence of Balmer H, FeI, MgI, NaI and HeI emission lines, which form on the side of the secondary-star facing the white dwarf. Their radial velocity curves are roughly consistent with each other but different from the secondary-star radial velocity curve measured by Howell et al. (2006a). Also, the application of the inverse K-correction is insufficient to produce agreement between our and the Howell et al. radial velocity.
 
The emission line fluxes show a bell shaped modulation across the orbit during epoch 1 observations, which is often taken as irradiation of the secondary-star. 
We produced irradiated secondary-star models tailored for VV~Pup and, though irradiation predicts low ionization metal lines, it does not explain the Balmer, Na and HeI emission lines strengths we observed. 
In particular, to produce these emission lines 
would require a $>$20000K white dwarf in VV Pup, contrary to the UV observations. The possible presence of a hot spot (20000 to 24000K) on the magnetic white dwarf, similar to that observed in low-state polars such as EF Eri and AM Her, could alternatively provide some Balmer, Na and possibly He emission lines. 
However, in epoch 2 observations taken while VV~Pup was experiencing an increase in mass transfer, we observed the same lines to strengthen and show a more scattered flux distribution with the orbital period. 
This also disfavors irradiation of the secondary-star. 

Our observations are better explained by a scenario in which the H, Na and HeI emission lines form mainly on  the white dwarf facing side of the secondary-star either in the threading region close to the L1 point, or in chromospheric prominence-loops possibly shocked near the location of the white dwarf magnetospheric radius. 
We note that the NaI doublet and  the H$\alpha$ emission  lines are known to be good indicators of chromosphere activity in M and L dwarfs (Andretta et  al.  1997, Schmidt et al.  2007). Flux ratios of the HeI$\lambda$5875/HeI$\lambda$6678$\sim$2.5-3.5, as measured by us in the ``bursting'' spectra of epoch 2, are  consistent with  low density gas regions  having temperatures  higher than 8000K  as in the stellar chromosphere or chromosphere-like regions (Giampapa  et al.   1978; Howell  et  al. 2006b). In addition, the strengthening of the chromospheric activity line indicators is consistent with stellar activity phenomena which, depending on the secondary-star dynamo magnetic field, might trigger episodes of enhanced mass transfer rate as observed in our epoch 2 ``bursting'' spectra.

\begin{acknowledgements}
EM  thanks Vincenzo Andretta  for his  help providing  valuable inputs
about chromosphere activity. EM also thanks the ESO director general for having supported a science leave in Bologna where the paper was concluded and the Bologna IASF/INAF institute for the kind hospitality. The authors wish also to thank the anonymous referee for the precious suggestions and comments while reviewing the first draft.  `
\end{acknowledgements}

\end{document}